\documentclass[prb,onecolumn,floatfix,notitlepage,showpacs,epsf,amssymb,longbibliography]{revtex4-1}
\usepackage{graphicx,amsmath,bm,colordvi,times}
\usepackage{multirow}
\usepackage{epstopdf}
\usepackage[english]{babel}
\usepackage{amsfonts}
\usepackage{amssymb}
\usepackage{float}
\usepackage{hyperref}
\usepackage[raggedright]{titlesec}
\hypersetup{pdfborder=0 0 0,colorlinks=true,citecolor=blue,linkcolor=blue}
\begin{document}

\title{Electron-hole superfluidity in strained Si/Ge type II heterojunctions}
\author{Sara Conti*$^{1}$, Samira Saberi Pouya$^1$,Andrea Perali$^2$, 
 Michele Virgilio$^3$, Fran\c{c}ois M. Peeters$^{1}$, Alexander R. Hamilton$^4$, Giordano Scappucci*$^5$ 
 \& David Neilson$^{1,4}$}
\affiliation{
$^1$ Department of Physics, University of Antwerp, Groenenborgerlaan 171, 2020 Antwerp, Belgium\\
$^2$ Supernano Laboratory, School of Pharmacy, Universit\`a di Camerino, 62032 Camerino (MC), Italy\\
$^3$ Department of Physics ``Enrico Fermi'', Universit\`a di Pisa, Largo Pontecorvo 3, 56127 Pisa, Italy\\
$^4$ ARC Centre of Excellence for Future Low Energy Electronics Technologies, School of Physics,
The University of New South Wales, Sydney, New South Wales 2052, Australia\\
$^5$ QuTech \& Kavli Institute of Nanoscience, Delft University of Technology, PO Box 5046, 2600GA Delft, The Netherlands\\
*email:Sara.Conti@uantwerpen.be,  G.Scappucci@tudelft.nl
}

\begin{abstract}
Excitons are promising candidates for generating superfluidity and Bose-Einstein Condensation (BEC) in  solid state devices, 
but an enabling material platform with in-built bandstructure advantages and scaling compatibility with industrial semiconductor technology is lacking.
Here we predict that spatially indirect excitons in a lattice-matched strained Si/Ge bilayer embedded into a germanium-rich SiGe crystal, 
would lead  to observable mass-imbalanced electron-hole superfluidity and BEC. 
Holes would be confined in a compressively strained Ge quantum well and electrons in a lattice-matched tensile strained Si quantum well. 
We envision a device architecture that does not require an insulating barrier at the Si/Ge interface, since this interface offers a type II band alignment.  
Thus the electrons and holes can be kept very close but strictly separate, strengthening the electron-hole pairing attraction while preventing fast electron-hole recombination. 
The band alignment also allows a one-step procedure for making independent contacts to the electron and hole layers, 
overcoming a significant obstacle to device fabrication. 
We predict superfluidity at experimentally accessible temperatures of a few Kelvin and carrier densities up to $\sim 6\times 10^{10}$ cm$\bm{^{-2}}$, 
while the large imbalance of the electron and hole effective masses can lead to exotic superfluid phases.
\end{abstract}

\maketitle

\section*{Introduction}
\vskip-\baselineskip

Spatially indirect excitons in a semiconductor system are a highly sought alternative for achieving quantum condensation and superfluidity in solid state devices at experimentally accessible temperatures.
Excitons are electrons and holes that bind into pairs through their long-range Coulombic attraction. 
Spatially indirect excitons have the electrons and holes confined in two separated but closely adjacent quantum wells or quasi two-dimensional (2D) layers\cite{Lozovik1975,Lozovik1976}. 
In the spatially indirect configuration, the electron-hole attraction can be very strong, 
while at the same time the electrons and holes are prevented from mutually annihilating through recombination.

From an application perspective, a supercurrent in the electron-hole superfluid could carry an electric current if the electron and hole layers are independently contacted in a counterflow configuration, 
directly leading to applications in dissipationless solid state electronics\cite{Su2008,Nandi2012}.  
Furthermore, the superfluid can be continuously tuned from the strongly coupled BEC bosonic regime to the BCS-BEC crossover regime of less strongly coupled fermionic pairs, simply by varying the carrier density using metal gates. 
In addition, when the electron and hole masses in a semiconductor are different, there are predictions of exotic superfluid phases\cite{Pieri2007}, 
including the Fulde-Ferrell-Larkin-Ovchinnikov (FFLO) phase\cite{Wang2017a} and the Sarma phase with two Fermi surfaces\cite{Forbes2005}.
These exotic phases are predicted to occur at much higher temperatures than in mass-imbalanced ultracold atomic gas Fermi mixtures\cite{Kinnunen2018, Frank2018}.
 
Two-dimensional van der Waals systems show particular promise because they offer the possibility of ultra-thin insulating barriers and conducting layers,
with very strong electron-hole pairing interactions as a result.
There have been predictions of quantum condensation of spatially indirect excitons in double bilayer graphene\cite{Perali2013} and double Transition Metal Dichalcogenide (TMD) monolayers\cite{Fogler2014}, 
and recent experiments have provided strong evidence for quantum condensation in these systems\cite{Burg2018,Wang2019}.  
One aspect of graphene bilayers and TMD monolayers is that the electron and hole effective masses are nearly equal, making these systems unsuitable for generating exotic superfluid phases.
However, the most pressing limitation of the 2D van der Waals systems lies in the rudimentary methods employed for device fabrication. 
These entail a layer by layer assembly using pick and transfer techniques which are prone to layer wrinkling, contamination and misorientation\cite{Frisenda2018},
leading to very poor device yield with limited prospects for scalability.

A more scalable approach is based on spatially indirect excitons in conventional semiconductor heterostructures, such as electrons and holes in GaAs double quantum wells (DQW)\cite{Croxall2008,Seamons2009}. 
However, despite indications of possible quantum condensation at very low temperatures (below $1$ K), 
concrete evidence for equilibrium BEC or superfluidity has remained elusive in this rather mature material system. 
It has been shown that the intrinsic properties of the GaAs/AlGaAs band structure constitute the main limitations for excitonic condensation\cite{SaberiPouya2020}, and also pose severe challenges in device fabrication and operation\cite{DasGupta2011a}.
First, the type I GaAs/AlGaAs band alignment makes it difficult to develop independent and selective contacts to the electron and hole layers. 
Second, in GaAs electron-hole DQWs, the energy separation between electron and hole states is $\approx 1.5$ eV, requiring rather wide AlGaAs barriers to avoid interlayer leakage.  
As a consequence, the electron-hole mutual Coulomb attraction is relatively weak and exciton formation greatly suppressed. 
Third, GaAs heterostructures are grown by molecular beam epitaxy. 
This growth technique is not compatible with conventional complementary metal-oxide semiconductor technology, so that prospects for advanced manufacturing and large scale device integration are severely limited. 

For these reasons, investigation of other solid-state systems that may overcome some limitations of double layers in GaAs, graphene and TMDs is of great interest.
In this letter we propose as a candidate for electron-hole superfluidity and BEC, an alternative mass-imbalanced solid-state system: 
a lattice-matched strained Si/Ge bilayer embedded into a germanium-rich SiGe crystal. 
Holes are confined in a compressively strained Ge quantum well and electrons in a tensile strained Si quantum well, with no barrier in between. 
This is possible since the Si/Ge interface offers a type II band alignment\cite{schaffler_high-mobility_1997,lee_strained_2004,virgilio_type-i_2006}, 
and thus electrons and holes can be kept separate but very close together. 
This enhances the strength of the electron-hole attraction while preventing unwanted recombination. 

This alternative route is promising since Si and Ge heterostructures have reached maturity in the past decade. 
Si and Ge heterostructures have very low disorder, with carrier mobilities exceeding one million in both constituents of the bilayer: 
the 2D electron gas in Si/SiGe\cite{lu_observation_2009} and the 2D hole gas in Ge/SiGe\cite{dobbie_ultra-high_2012}. 
Furthermore, the carrier density may be tuned over orders of magnitude by leveraging on industrial gate-stack technology\cite{wuetz_multiplexed_2020,lodari_low_2020}. 
This material system also integrates with advanced quantum technologies\cite{Vandersypen2017InterfacingCoherent,scappucci_germanium_2020}, 
including long-lived electron spin qubits in Si\cite{Yoneda2018A99.9,watson_programmable_2018}, 
hole spin qubit arrays in Ge\cite{hendrickx_four-qubit_2020,hendrickx_fast_2020}, and 
superconducting contacts to holes\cite{hendrickx_gate-controlled_2018,hendrickx_ballistic_2019,vigneau_germanium_2019}.
A major advantage over the GaAs and TMD material systems is that the band alignments of the proposed Si/Ge bilayer should allow electrons in Si and holes in Ge to be contacted independently and selectively using a one step fabrication process. 
Due to strain, there is a very large mass imbalance between holes in Ge ($0.05 m_e$)\cite{lodari_light_2019} and electrons in Si ($0.2 m_e$)\cite{schaffler_high-mobility_1997,Zwanenburg2013}, opening the door to exploration of exotic superfluid phases.
Finally, these Si/Ge heterostructures may be grown on $300$ mm Si wafers using mainstream chemical vapor deposition\cite{Pillarisetty2011Academic}, and may profit from advanced semiconductor manufacturing for high device yield and integration.

Our calculations show that the envisaged Si/Ge bilayer supports a superfluid condensate. 
Tuning the carrier density continuously sweeps the superfluid across the BEC and BCS-BEC regimes of the superfluid, with an accompanying variation in the magnitude of the superfluid gap and the transition temperature.

\section*{Results}
\vskip-\baselineskip

\subsection*{Material stack and device architecture}
\vskip-\baselineskip

\begin{figure*}
\begin{center}
\includegraphics[angle=0,width=0.92\textwidth] {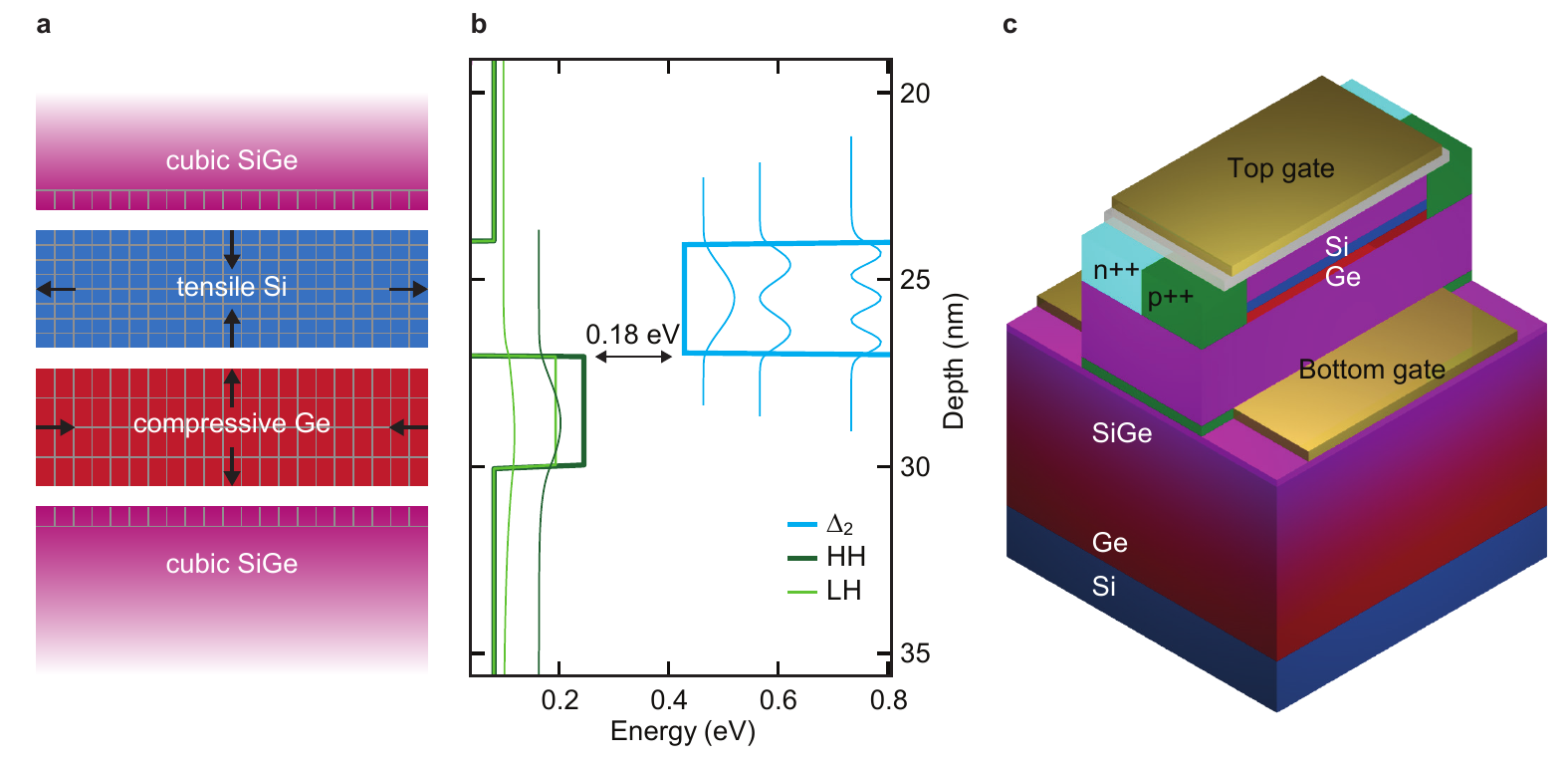}
\end{center}
\caption{
\textbf{Lattice-matched double Si/Ge bilayer, bandstructure, device architecture.}
\textbf{a} Below the critical thickness for plastic relaxation, epitaxial Ge and Si layers are lattice matched to the underlying 
SiGe, with compressive and tensile strain, respectively.
\textbf{b} Band structure of low-lying bands resulting in a double quantum well structure. 
Due to the type II band alignment at the strained-Si/strained-Ge heterojunction, electrons ($\Delta_2$ states) are confined in tensile strained Si and holes (HH states) in compressively strained Ge.
\textbf{c} Material stack with embedded Si/Ge bilayer and an envisaged device architecture, characterized by independent n$^{++}$ and p$^{++}$ contacts and gate electrodes to tune independently the electron and hole densities in the bilayer.
} 
\label{Fig:Materials}
\end{figure*}

Figure \ref{Fig:Materials}a illustrates the concept of a lattice-matched Si/Ge bilayer. 
A cubic (strain-relaxed) Ge-rich Si$_{1-x}$Ge$_x$ substrate, with a Ge concentration $x=0.8$, sets the overall in-plane lattice parameter of the stack. 
For layer thicknesses below the critical thickness for onset of plastic relaxation, the Si/Ge epilayers will grow with the same in-plane lattice constants as the underlying Si$_{0.2}$Ge$_{0.8}$\cite{matthews_defects_1976,Paul2010}.
Therefore, the Ge layer will be compressively strained in the in-plane direction\cite{sammak_shallow_2019} and, conversely, 
the Si layer will be under tensile strain\cite{schaffler_high-mobility_1997}. 
A thickness of $3$ nm for each Ge and Si layer allows for such strain engineering\cite{Paul2010}, and is feasible experimentally given the recent advances in low-temperature chemical vapor deposition of SiGe heterostructures comprising Si and Ge quantum wells\cite{dyck_accurate_2017,sammak_shallow_2019}. 

Figure \ref{Fig:Materials}b shows the band-structure of the proposed lattice matched Si/Ge bilayer, where the Ge and Si layers are strained in opposite directions. 
The band-structure was calculated in an effective mass approach using deformation potential theory to take into account the impact of the strain field on the relevant band edges. 
We highlight three highly attractive features. 
First, there is effectively only one quantum well for electrons, spatially separated from the single quantum well for holes.
There does exist a quantum well for electrons in Ge\cite{virgilio_type-i_2006}, but it is located so high in energy it would remain inactive for the present phenomenon.
This is quite different from GaAs and TMD's, where each of the two layers has both electron and hole wells.   
The valence band profile shows that the wave-function of the fundamental heavy hole (HH) state is confined in the Ge quantum well. 
The large energy splittings between the fundamental HH state and the fundamental light hole (LH) state are the result of the compressive strain and the larger confinement mass of the HH with respect to the LH.   
The conduction band profile shows that the wave-function of the fundamental $\Delta_2$ state is confined in the Si quantum well, with other states being higher up in energy. 
There is no hole quantum well in the Si layer.
The second feature is that the energy difference between the bottom of the conduction band and the top of the valence band is $\approx0.18$ eV. 
This is $\sim 8$ times smaller than in GaAs quantum wells, meaning a small interlayer bias should be sufficient for tuning of the electron and hole wave-function shape and position in the quantum well.
Finally, based on previous theoretical predictions and experiments, such strained Ge and Si quantum wells will result in a large imbalance of the masses, with a very light in-plane effective mass for holes ($\approx0.05m_e$) and a much heavier in-plane effective mass for electrons ($\approx0.19m_e$).
This has important implications for the superfluid, as detailed in the subsection {\it Screening polarizabilities in the superfluid state}.

Figure \ref{Fig:Materials}c illustrates the entire material stack with the embedded Si/Ge bilayer and an envisaged device architecture. 
Starting from a Si(001) wafer and a thick strain-relaxed Ge layer deposited on top, 
a high-quality Si$_{0.2}$Ge$_{0.8}$ strain-relaxed buffer is obtained by reverse grading the Ge content in the alloy\cite{shah_reverse_2008,sammak_shallow_2019}. 
Importantly, the strain-relaxed Si$_{0.2}$Ge$_{0.8}$ buffer can be heavily $p$-type doped to serve as an epitaxial bottom gate.
This bottom gate can be biased negatively to populate only the undoped Ge quantum well with holes. 
Following the deposition of the Si/Ge bilayer, an additional Si$_{0.2}$Ge$_{0.8}$ barrier separates the Si quantum well from a dielectric layer and top gate. 
The top gate may be biased positively to populate only the undoped Si quantum well with electrons. 
Separate and independent Ohmic contacts to the capacitively induced 2D electron gas in Si and hole gas in Ge is achieved by standard n$^{++}$ and p$^{++}$ ion implantation, respectively\cite{lee_strained_2004,Pillarisetty2011Academic}. 
Alternatively, the p$^{++}$ contact may be substituted with an aluminium layer or an in-diffused metallic germano-silicide, as routinely done in Ge/SiGe heterostructure field effect transistors\cite{sammak_shallow_2019}.
By carefully designing the thickness of the Si$_{0.2}$Ge$_{0.8}$ barriers above and below the Si/Ge bilayer, we envisage that the carrier density may be tuned independently in the electron and hole layer in the low density regime, $n< 10^{11}$ cm$^{-2}$. 

Independent electrical contact to the two layers is easily achieved thanks to the remarkable, unique band structure of the material stack, with only one quantum well for electrons in the strained Si and only one accessible quantum well for holes in the strained Ge. 
Crucially, the prospect to make independent contacts to the electron-hole bilayer with a simple and robust process, in contrast to the challenging processing required for selective contacts in III-V and TMD materials, bodes well for superfluidity measurements in counterflow configurations\cite{su_how_2008,DasGupta2011a}.

\subsection*{Superfluid properties}
\vskip-\baselineskip

\begin{figure}[h]
\begin{center}
\includegraphics[width=0.56\textwidth] {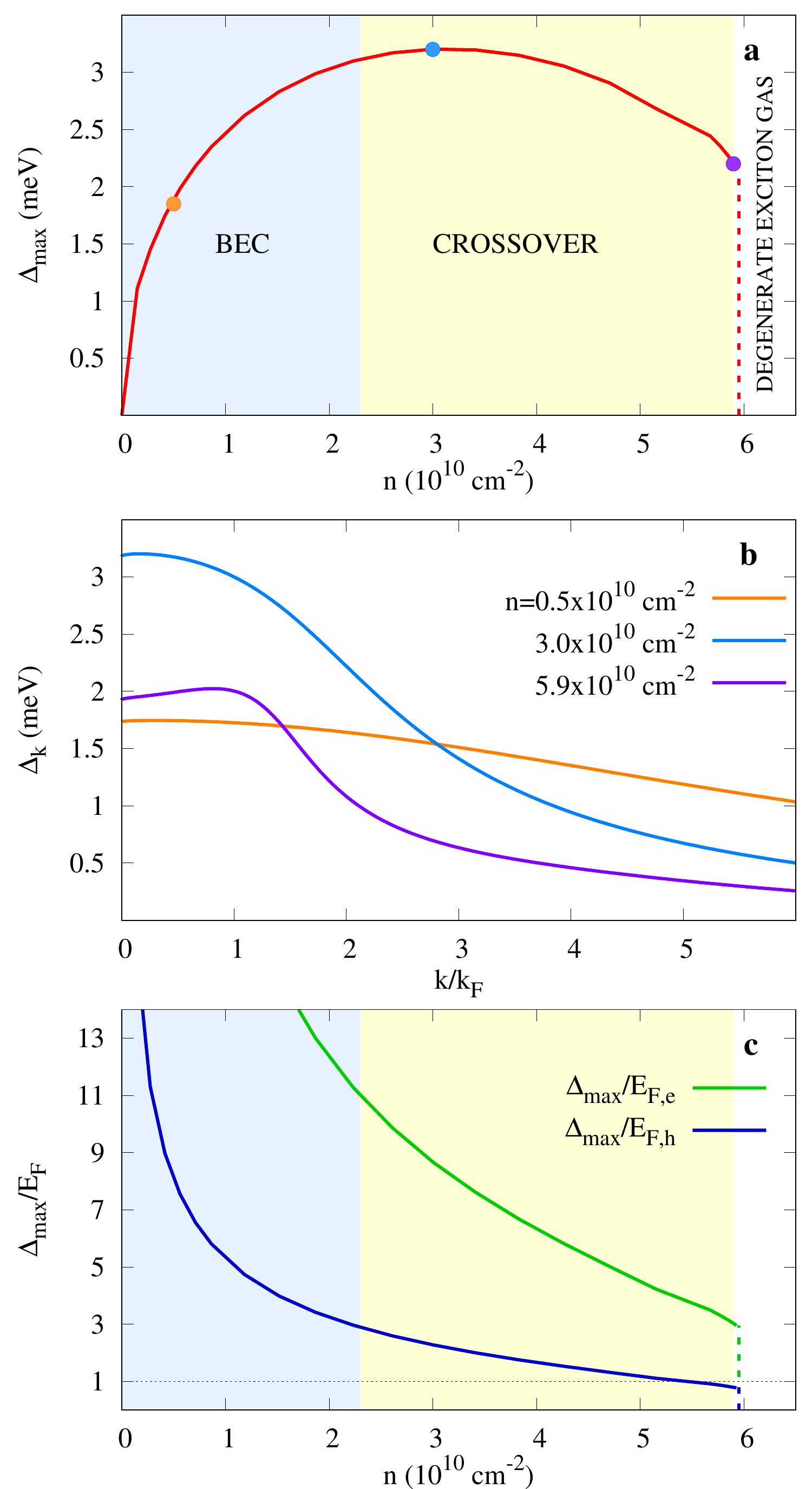}
\end{center}
\caption{
\textbf{Properties of the zero-temperature superfluid gap.}  
\textbf{a} Maximum of the gap $\Delta_{\mathrm{max}}$ as a function of the equal carrier densities $n$. 
The blue and  yellow areas represents the BEC and BCS-BEC crossover regimes. 
The BCS regime is suppressed by screening.
\textbf{b} Momentum dependent gap $\Delta_{\mathbf{k}}$ at the three densities $n$ indicated by the colour-coded dots on the curve in \textbf{a}. 
\textbf{c} $\Delta_{\mathrm{max}}$ scaled by the electron and hole Fermi energies $E_{F,e}$ and $E_{F,h}$.  
}
\label{Fig:SFGap}
\end{figure}

In Fig.\ \ref{Fig:SFGap} we report superfluid properties of the system, calculated including self-consistent screening 
of the electron-hole pairing interaction.  
Self-consistent treatment of the long-ranged Coulombic pairing interaction\cite{Lozovik2012} is a key element for determining electron-hole superfluid properties,  
in contrast to other superconductors and superfluids with short-range pairing interactions.  

Figure \ref{Fig:SFGap}a shows the maximum of the zero-temperature superfluid gap $\Delta_{\mathrm{max}}$ as a function of the electron and hole densities, assumed equal.
As the density $n$ increases, $\Delta_{\mathrm{max}}$ first increases, passes through a maximum and then decreases.  
Above an onset density, $n_0\sim 5.9 \times 10^{10}$ cm$^{-2}$, $\Delta_{\mathrm{max}}$ very rapidly drops to negligible values, $\lesssim 1$ $\mu$eV. 
The behaviour of $\Delta_{\mathrm{max}}$ is a consequence of the self-consistency included in the screening\cite{Perali2013}.
This has radically different effects in the different regimes of the superfluidity\cite{Neilson2014,LopezRios2018}.   
(i) At low carrier densities, the binding energy of the electron-hole pairs is large relative to the Fermi energy, and the excitons are strongly bound and compact. 
They resemble weakly-interacting, approximately neutral, bosons and screening is negligible.
(ii) With increasing density, the number of pairs increases and the gap gets stronger.  
However, relative to the Fermi energy the gap gets weaker, and the superfluid moves from the BEC regime of bosons to the less strongly coupled BCS-BEC crossover regime of coupled fermionic pairs.  
(iii) When the density is further increased above the onset density, $n>n_0$, strong screening overcomes the weak electron-hole pair coupling in what would be the BCS regime, and the superfluidity is suppressed.

The density ranges for the BEC and BCS-BEC crossover regimes are indicated in Fig.\ \ref{Fig:SFGap}a by the blue and yellow shaded areas. 
We characterize the regimes by the condensate fraction parameter $c$, the fraction of carriers bound in pairs\cite{Salasnich2005,Guidini2014}.  
The boundary between the BEC and BCS-BEC crossover regimes is determined by $c = 0.8$.  
The BCS-BEC crossover to BCS boundary would be located at $c=0.2$, but at the onset density $n_0$ the condensate fraction has only dropped to $c \sim 0.5$, implying the absence of a BCS regime.  

Figure \ref{Fig:SFGap}b shows the momentum dependence of the superfluid gap $\Delta_{\mathbf{k}}$ at three densities $n$.  
Near the onset density $n_0=5.9 \times 10^{10}$ cm$^{-2}$, which is in the BCS-BEC crossover regime, $\Delta_{\mathbf{k}}$ has a broad peak centred close to $k=k_F$, indicating proximity of the BCS regime.  
At density $n=3.0 \times 10^{10}$ cm$^{-2}$ corresponding to the maximum of $\Delta_{\mathrm{max}}$, which is near the boundary separating the BCS-BEC crossover and BEC regimes, the peak in $\Delta_{\mathbf{k}}$ has moved back to ${\mathbf{k}}=0$.  
At $n=0.5 \times 10^{10}$ cm$^{-2}$, which is in the deep BEC regime, $\Delta_{\mathbf{k}}$ extends out to large $k/k_F$.  

Figure \ref{Fig:SFGap}c shows $\Delta_{\mathrm{max}}$ scaled to the electron and hole Fermi energies. 
At the onset density, while $\Delta_{\mathrm{max}}<E_{F,h}$, it is surprising that $\Delta_{\mathrm{max}} \gg E_{F,e}$.  
This result significantly differs from the equal mass case, for which the onset density occurs around  
$\Delta_{\mathrm{max}}/E_{F,h}=\Delta_{\mathrm{max}}/E_{F,e}\sim 1$.
We recall the physical argument that a sufficiently strong energy gap $\Delta_{\mathrm{max}}$ relative to the Fermi energy, will exclude from the screening low-lying excited states that otherwise would very significantly weaken the electron-hole attraction\cite{Neilson2014}.  
But it is puzzling why $\Delta_{\mathrm{max}}$ must become so much larger than $E_{F,e}$ before the screening is suppressed by the superfluidity.   
To explain this, we must look at the self-consistent screening polarizabilities in the superfluid state when the electron and hole masses are unequal.

\subsection*{Screening polarizabilities in the superfluid state}
\vskip-\baselineskip

\begin{figure}[!h]
\begin{center}
\includegraphics[width=0.52\textwidth] {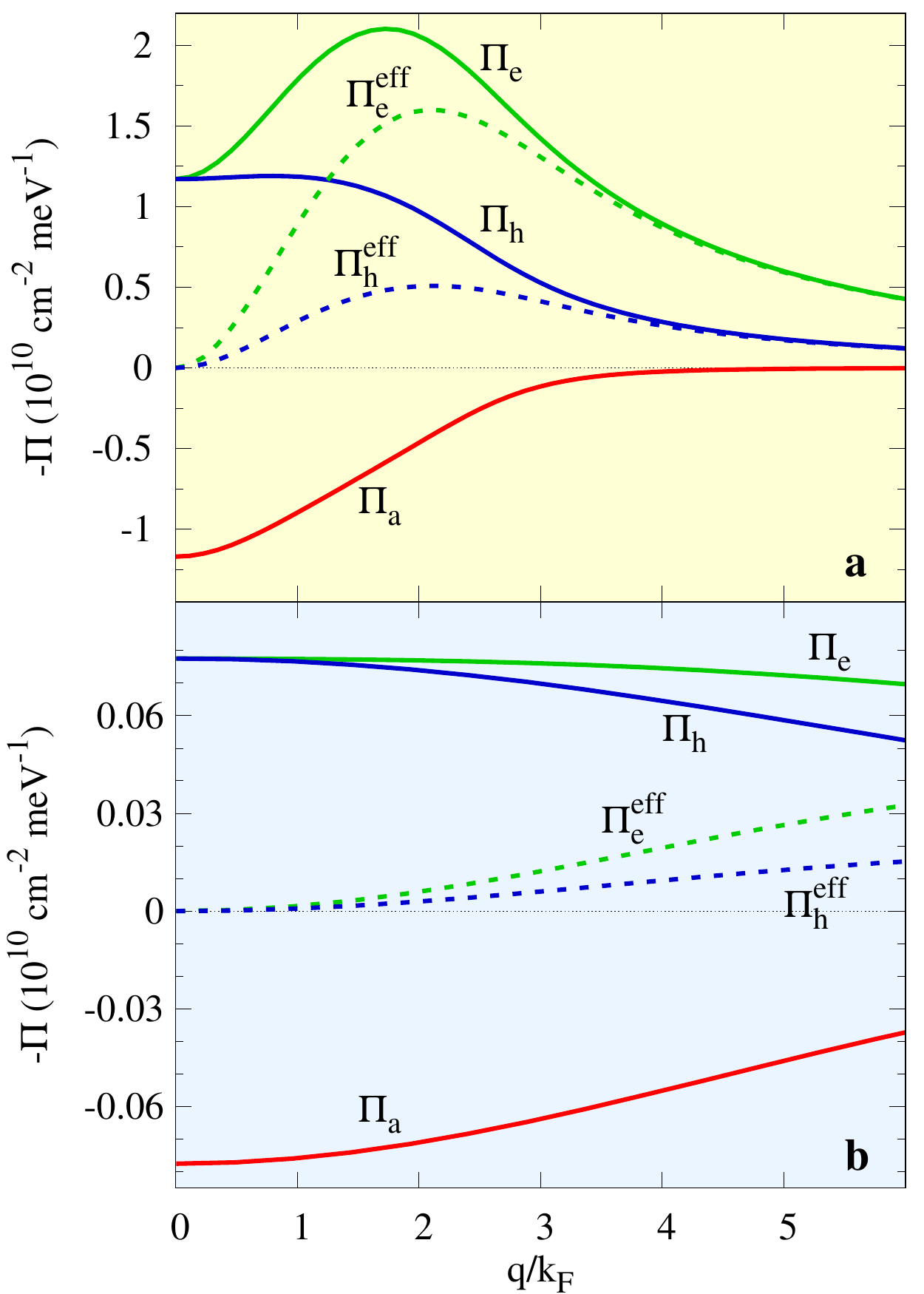}
\end{center}
\caption{
\textbf{Polarizabilities.}  
The green (blue) solid lines are the normal electron (hole) polarizabilities in the presence of the superfluid $\Pi_e(\mathbf{q})$ ($\Pi_h(\mathbf{q})$).  
The red solid lines are the anomalous polarizabilities $\Pi_a(\mathbf{q})$ for the superfluid electron-hole pairs.  
Dashed green and blue lines are the corresponding effective polarizabilities  $\Pi^{\rm{eff}}_{e,h}(\mathbf{q}) \equiv \Pi_{e,h}(\mathbf{q}) +\Pi_a(\mathbf{q})$.  
{\bf a}\ \ For density $n\simeq n_0=5.9 \times 10^{10}$ cm$^{-2}$, in the BCS-BEC crossover regime. 
{\bf b}\ \ For density $n=0.5 \times 10^{10}$ cm$^{-2}$, in the 
BEC regime.
}
\label{Fig:Pol}
\end{figure}

We now examine the effect of the different masses on the polarizabilities that control the self-consistent screening.
Figure \ref{Fig:Pol} shows the normal polarizabilities for electrons and holes in the presence of the superfluid, 
$\Pi_e(\mathbf{q})$ and $\Pi_h(\mathbf{q})$ (Eqs.\  (\ref{Eq:Pi_e}) and (\ref{Eq:Pi_h})), 
and the anomalous polarizability for the superfluid electron-hole pairs, $\Pi_a(\mathbf{q})$ (Eq.\ (\ref{Eq:Pi_a})). 
The panels are for two densities, the first close to the onset density in the BCS-BEC crossover regime and the second in the BEC regime. 

If the effect of the superfluidity on the screening is not taken into account, the polarizabilities would be much larger than $\Pi_e(\mathbf{q})$ and $\Pi_h(\mathbf{q})$.  
In 2D they would be given in the momentum transfer range relevant for screening $\mathbf{q}/k_F\leq 2$, by their respective densities of states, 
\begin{align}
\Pi_e^{(N)}(\mathbf{q}) &=\frac{m^\star_e}{\pi\hbar^2} \simeq 8.0\times 10^{10}\text{cm}^{-2} \text{meV}^{-1}  \nonumber \\
\Pi_h^{(N)}(\mathbf{q}) &=\frac{m^\star_h}{\pi\hbar^2} \simeq 2.1\times 10^{10}\text{cm}^{-2} \text{meV}^{-1}  	\ . 
\label{Eq:Pi^(N)} 
\end{align}
Such large polarizabilities would lead to very strong screening of the electron-hole interaction, resulting in extremely weak-coupled superfluidity and very low transition temperatures in the mK range\cite{Neilson2014}.
The suppression of the normal polarizabilities in the superfluid state arises from the blocking by the superfluid gap of low-lying states in the energy spectrum (Eqs.\ (\ref{Eq:Pi_e}) and (\ref{Eq:Pi_h})).
There is even more suppression of screening that comes from cancellation with the anomalous polarizability in the screened interaction (Eq.\ (\ref{Eq:VeffSF})).  
To highlight these cancellations, in Fig.\ \ref{Fig:Pol} we introduce effective polarizabilities,  
$\Pi^{\rm{eff}}_e(\mathbf{q}) = \Pi_e(\mathbf{q}) +\Pi_a(\mathbf{q})$ and
$\Pi^{\rm{eff}}_h(\mathbf{q}) = \Pi_h(\mathbf{q})+\Pi_a(\mathbf{q})$. 

At $\mathbf{q}\!=\!0$ and for all densities, there is the property
$\Pi_e(\!\mathbf{q}\!=\!0\!)=\Pi_h(\!\mathbf{q}\!=\!0\!)=-\Pi_a(\!\mathbf{q}\!=\!0\!)$, 
so that $\Pi^{\rm{eff}}_e(\!\mathbf{q}\!=\!0\!) = \Pi^{\rm{eff}}_h(\!\mathbf{q}\!=\!0\!)=\!0$,  
reflecting the absence of long distance screening for any non-zero superfluid gap.  
For non-zero $\mathbf{q}$, however, the cancellation of polarizabilities and the suppression of screening are very sensitive to which regime the superfluidity lies in.
$\Delta_{\mathbf{k}}$ becomes narrower as we move from the BEC regime, across the BCS-BEC crossover regime, and towards the BCS regime (Fig.\ \ref{Fig:SFGap}b), narrowing both the momentum range for blocked excitations and the momentum range for which the cancellations are significant.

Figure \ref{Fig:Pol}a is in the BCS-BEC crossover regime, where we see that the behaviour of $\Pi_e(\mathbf{q})$ for non-zero $\mathbf{q}$ is strikingly different from $\Pi_h(\mathbf{q})$. 
We discuss details of their differences in functional behaviour in the Supplementary Material.  
We also note the approximate cancellation of $\Pi_a(\mathbf{q})$ with $\Pi_h(\mathbf{q})$ but not with $\Pi_e(\mathbf{q})$. 
This is because $\Pi_a(\mathbf{q})$ depends on the strength of the pairing, and so scales with the reduced mass $m^\star_r$, which for this system is approximately equal to $m^\star_h$. 
In contrast $m^\star_r \ll m^\star_e$, so $\Pi_e(\mathbf{q})$ is larger than $\Pi_a(\mathbf{q})$ and does not cancel with it. 
We conclude for intermediate values of $k/k_F$, that  $\Delta_{\mathbf{k}}$ strongly suppresses the effective polarization for holes, 
$\Pi^{\rm{eff}}_h(\mathbf{q})$, but does not suppress the effective polarization for electrons, $\Pi^{\rm{eff}}_e(\mathbf{q})$. 
This explains the puzzling result we noted in Fig. \ref{Fig:SFGap}c, that the screening is only suppressed when the gap reaches very large values of $\Delta_{\mathrm{max}}/E_{F,e} \gg 1$. 

In contrast, Fig.\ \ref{Fig:Pol}b shows in the deep BEC regime for $\mathbf{q}/k_F\lesssim 2$, that 
$\Pi_e(\mathbf{q})$ and $\Pi_h(\mathbf{q})$ are very similar and now both scale with the reduced mass $m^\star_r$, like $\Pi_a(\mathbf{q})$.  
This is because in this regime the electrons and holes are in strongly bound pairs that have lost their single particle character. 
As a result, $\Pi^{\rm{eff}}_e(\mathbf{q})$ and $\Pi^{\rm{eff}}_h(\mathbf{q})$ are very small over the momentum transfer range important for screening, reflecting near complete cancellation.
Physically, the electron-hole pairs in the deep BEC are compact compared with the inter-particle spacing and approximately neutral, and this makes screening unimportant.  

\subsection*{Superfluid phase diagram}
\begin{figure}[h]
\begin{center}
\includegraphics[width=0.5\textwidth]{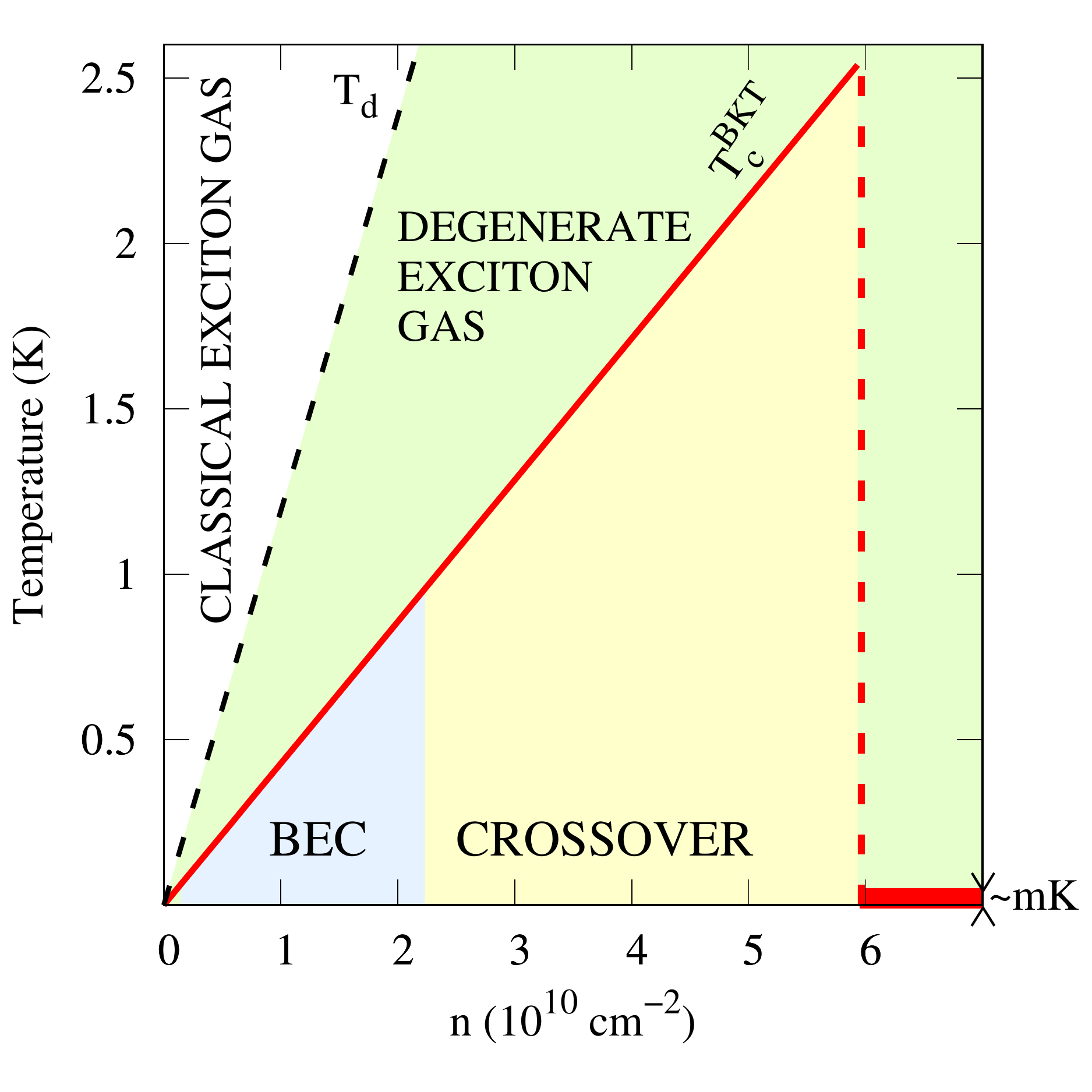}
\caption{\textbf{Phase diagram.}  
$T_c^{BKT}$ is the superfluid transition temperature.  
$n$ is the density of the electrons and holes.
$T_d$ is the degeneracy temperature for the exciton gas.
The superfluid BEC and BCS-BEC crossover regimes are indicated. 
Above $n\sim 6\times 10^{10}$ cm$^{-2}$, the superfluid transition temperature is in the mK range.}
\label{fig:PhaseDia}
\end{center}
\end{figure}

\vskip-\baselineskip
The superfluid transition for a 2D system is a Berezinskii-Kosterlitz-Thouless (BKT) transition\cite{Kosterlitz1973}.  
For parabolic bands, the transition temperature $T_c^{BKT}$ becomes linearly proportional to the carrier density $n$ (Eq.\ (\ref{T_KT_n})).  
The highest transition temperature occurs at the onset density. 
The complete phase diagram is shown in Fig.\ \ref{fig:PhaseDia}. 
Despite the large dielectric constant and the small hole mass, we see that the transition temperatures are readily experimentally accessible, up to $T_c^{BKT}= 2.5$ K.  

At $T_c^{BKT}$, there is a thermally driven transition to a degenerate exciton gas, in which the system has lost its macroscopic coherence, but  local pockets of superfluidity remain.   
These pockets persist up to a characteristic degeneracy temperature $T_d$\cite{Butov2004}.   
At $T_d$, the excitons lose degeneracy and the system becomes a classical exciton gas. 

When the density is increased to an onset density $n_0=5.9\times 10^{10}$ cm$^{-2}$, the superfluid gap drops nearly discontinuously to exponentially small values (Fig.\ \ref{Fig:SFGap}a), and $T_c^{BKT}$ drops to the sub-mK range.  
The drop in the gap is similar to a first order transition, and is caused by the sudden collapse of three solutions to the gap equation (Eq.\ (\ref{Eq:Delta})) into a single very low-energy solution\cite{Lozovik2012}. 
This is caused by strong screening of the electron-hole pairing attraction. 

\section*{Discussion}
\vskip-\baselineskip
In summary, we predict superfluidity in experimentally accessible samples, densities and temperatures, with a superfluid gap of up to $3$~meV and a transition temperature up to $2.5$~K. 
At carrier densities higher than $6\times 10^{10}$ cm$^{-2}$, the relatively weak superfluidity is unable to suppress the increasingly strong screening, with a result that screening suppresses superfluidity. 
The nature of this self-consistent screening process is sensitive to the system property that the electron and hole masses are markedly different. 
Since the existence of exotic superfluid phases is dependent on unequal masses, 
and because the effect of unequal masses does not manifest itself in the BEC regime, ($n< 2\times 10^{10}$ cm$^{-2}$),
the experimental search for these exotic phases should focus on intermediate densities in the BCS-BEC crossover regime ($2\times 10^{10} < n < 6\times 10^{10}$ cm$^{-2}$).

In a realistic implementation of the proposed Si/Ge bilayer material stack, the Si/Ge interface will have a finite width due to segregation, diffusion, and intermixing associated with the chemical vapor deposition. 
However, by depositing the Si/Ge layers at sufficiently low temperatures ($\leq 500$~$^\circ$C as in Refs.~\onlinecite{sammak_shallow_2019,dyck_accurate_2017}), we foresee a distance for transitioning between the strained Ge and Si layer of much less than $1$~nm. 
The details of such Si/Ge transition region do not influence the main findings of the bandstructure calculations nor the superfluid properties, since the principle parameter affecting these is the distance separating the centres of the two wells. 

The superfluid gap shown in Fig.\ \ref{Fig:SFGap}a reaches $35$ K ($3$ meV), and the transition temperature could be increased up to a significant fraction of this value by implementing variations of the proposed Si/Ge material stack, 
for example by including a finite stack of bilayers that eliminates the limitations of a 2D system\cite{VanderDonck2020}.
Such Si/Ge material stacks with up to five bilayers are within reach, considering the recent experimental progress in the deposition of multi-quantum wells on Ge-rich SiGe virtual substrates\cite{grange_atomic-scale_2020}. 
In this way, the transition temperature would allow for the design and development of an entire new class of dissipationless logic devices and electronics\cite{reddy_bilayer_2009} for CMOS-based cryogenic control of silicon quantum circuits\cite{xue_cmos-based_2020}.

\section*{Methods}
\vskip-\baselineskip

For calculations of the superfluid properties, we take a structure with hole effective mass $m^\star_h=0.05m_e$ in the compressively strained Ge layer, and electron effective mass $m^\star_h=0.19m_e$ in the tensile strained Si layer. 
We use a uniform dielectric constant determined for Si and Ge quantum wells of equal width in contact, 
$\epsilon = 2\left(1/\epsilon_{Ge}+1/\epsilon_{Si}\right)^{-1}= 13.7$,
with $\epsilon_{Ge}=16.2$ and $\epsilon_{Si}=11.9$.
Lengths are expressed in units of the effective Bohr radius, 
$a^\star_{B} = \hbar^2 4\pi \epsilon_0\epsilon /({m^\star_r}e^2)=18.3$ nm, 
and energies in effective Rydbergs, $Ry^{\star}=e^2/(2a_{B}^{\star})=33$ K.
$m^\star_r$ is the reduced effective mass. 

\subsection*{Mean field equations}
\vskip-\baselineskip
Because of the different masses, there are distinct electron and hole normal Matsubara Green functions, 
\begin{align}
&G_{e}(\mathbf{k},\tau)=-\left\langle T_\tau\, c(\mathbf{k},\tau) c^\dagger(\mathbf{k},0)\right\rangle\, , \nonumber \\
&G_{h}(\mathbf{k},\tau)=-\left\langle T_\tau\, d(\mathbf{k},\tau) d^\dagger(\mathbf{k},0)\right\rangle\, .
\label{Eq:normalGreenFunctions} 
\end{align}  
$T_\tau$ is the ordering operator in imaginary time $\tau$. 
$c^{\dagger}$ and $c$ ($d^{\dagger}$ and $d$) are the creation and destruction operators for electrons (holes) in their respective quantum wells. 
The corresponding anomalous Green function is, 
\begin{equation}  
F(\mathbf{k},\tau)=-\left\langle T_\tau\, c(\mathbf{k},\tau) d(\mathbf{k},0)\right\rangle\, .
\label{Eq:anomalousGreenFunctions} 
\end{equation}  

In the weak-coupling BCS limit, Eqs.\  (\ref{Eq:normalGreenFunctions}) -- (\ref{Eq:anomalousGreenFunctions}) reduce to,
\begin{align}
G_{e}(i\omega_n, \mathbf{k})&=\frac{u_\mathbf{k}^2}{(i\omega_n-E^-_\mathbf{k})}+\frac{v_\mathbf{k}^2}{(i\omega_n+E^+_\mathbf{k})}\, ,\nonumber\\
G_{h}(i\omega_n, \mathbf{k})&=\frac{v_\mathbf{k}^2}{(i\omega_n+E^-_\mathbf{k})}+\frac{u_\mathbf{k}^2}{(i\omega_n-E^+_\mathbf{k})}\, ,\nonumber\\
F(i\omega_n, \mathbf{k})&=\frac{u_\mathbf{k}v_\mathbf{k}}{(i\omega_n-E^-_\mathbf{k})}-\frac{u_\mathbf{k} v_\mathbf{k}}{(i\omega_n+E^+_\mathbf{k})}\, ,
\label{Eq:GreenFunctions}
\end{align} 
where $\omega_n\ (n=1,2,3\dots)$ are the fermionic Matsubara frequencies and
%
\begin{equation}
E^{\pm}_{\mathbf{k}} = E_{\mathbf{k}} \pm \delta \xi_{\mathbf{k}}\, ,\qquad
E_{\mathbf{k}} = \sqrt{\xi_{\mathbf{k}}^2 + \Delta_{\mathbf{k}}^{2}}\, ,\qquad
\delta \xi_{\mathbf{k}} =\frac{1}{2}\left(\xi^h_{\mathbf{k}} - \xi^e_{\mathbf{k}}\right)\, ,\qquad 
\xi_{\mathbf{k}}=\frac{1}{2}\left(\xi^e_{\mathbf{k}} + \xi^h_{\mathbf{k}}\right).
\label{Eq:EnergyTerms}
\end{equation}
%
$\xi^e_{\mathbf{k}}= \frac{{k}^{2}}{2 m_{e}^\star}- \mu$ ($\xi^h_{\mathbf{k}}= \frac{{k}^{2}}{2 m_{h}^\star}- \mu$) is the electron (hole) single-particle energy band dispersion in the normal state, with $\mu$ the chemical potential.
$\Delta_{\mathbf{k}}$ is the superfluid energy gap.  
The Bogoliubov amplitudes are, 
$u_{\mathbf{k}}^2 = \frac{1}{2} \left(1+\frac{\xi_{\mathbf{k}}}{E_{\mathbf{k}}}\right)$ and 
$v_{\mathbf{k}}^2 = \frac{1}{2} \left(1-\frac{\xi_{\mathbf{k}}}{E_{\mathbf{k}}}\right)$.

We consider only equal electron and hole densities $n$.  
At zero temperature, the superfluid energy gap can be determined from the usual mean-field equation of BCS theory, even in the strongly interacting BCS-BEC crossover and BEC regimes:
\begin{equation}
\Delta_{\mathbf{k}}=\frac{1}{L^2}\!\sum_{\mathbf{k}',\omega_n}\!\!V^{sc}_{\mathbf{k}-\mathbf{k}'} F(i\omega_n, k')\!=
\!-\frac{1}{L^2}\!\sum_{\mathbf{k}'}\!V^{sc}_{\mathbf{k}-\mathbf{k}'} \frac{\Delta_{\mathbf{k}'} }{2 E_{\mathbf{k}'}}\ ,
\label{Eq:Delta}
\end{equation}
where $V^{sc}_{\mathbf{k}-\mathbf{k}'}=V^{sc}_{\mathbf{q}}$ is the attractive screened electron-hole interaction.  
As expected, the only mass parameter entering in Eq.\ (\ref{Eq:Delta}) is the reduced mass $m^\star_r$.
Equation (\ref{Eq:Delta}) is self-consistently solved coupled to the density equation,
\begin{equation}
n= \frac{2}{L^2} \sum_{\mathbf{k},\omega_n} G_\ell(i\omega_n, k)= \frac{2}{L^2} \sum_{\mathbf{k}} v_{\mathbf{k}}^2  \qquad \ell=e,h\ . 
\label{Eq:density}
\end{equation}  
For given density $n$, Eq.\ (\ref{Eq:density}) determines the chemical potential $\mu$.

\subsection*{Self-consistent screening}
\vskip-\baselineskip

Because the electron-hole interaction is Coulombic and long-ranged, it is essential to include screening in $V^{sc}_{\mathbf{q}}$.    
To determine the screening in the presence of a superfluid, we evaluate the density response functions within the Random Phase Approximation (RPA) for the double quantum well system in which the electrons and holes have different masses\cite{SaberiPouya2020}.
For the polarization loops, we use the normal and anomalous Green’s functions, Eqs.\  (\ref{Eq:GreenFunctions}).  
Then the normal polarizabilities in the presence of the superfluid are, 
\begin{align}
\Pi_e(\mathbf{q})&=
\frac{2}{L^2} \sum_{\mathbf{k}}\sum_{\omega_n} G_e(i\omega_n, \mathbf{k})G_e(i\omega_n, \mathbf{k}-\mathbf{q})\nonumber\\
&=-\frac{2}{L^2} \sum_{\mathbf{k}} 
\left[\frac{u^2_{\mathbf{k}}v^2_{\mathbf{k}-\mathbf{q}}}{E^{+}_{\mathbf{k}-\mathbf{q}}+E^-_{\mathbf{k}}}+
\frac{v^2_{\mathbf{k}}u^2_{\mathbf{k}-\mathbf{q}}}{E^{-}_{\mathbf{k}-\mathbf{q}}+E^{+}_{\mathbf{k}}}\right]\, ,
\label{Eq:Pi_e}
\end{align}
\begin{align}
\Pi_h(\mathbf{q})&=
\frac{2}{L^2} \sum_{\mathbf{k}}\sum_{\omega_n} G_h(i\omega_n, \mathbf{k})G_h(i\omega_n, \mathbf{k}-\mathbf{q})\nonumber\\
&=-\frac{2}{L^2} \sum_{\mathbf{k}} 
\left[\frac{u^2_{\mathbf{k}}v^2_{\mathbf{k}-\mathbf{q}}}{E^{-}_{\mathbf{k}-\mathbf{q}}
+E^+_{\mathbf{k}}}+\frac{v^2_{\mathbf{k}}u^2_{\mathbf{k}-\mathbf{q}}}{E^{+}_{\mathbf{k}-\mathbf{q}}+E^{-}_{\mathbf{k}}} \right]\ . 
\label{Eq:Pi_h}
\end{align}
The anomalous polarizability for the density response of the superfluid electron-hole pairs is,
\begin{align}	
\Pi_{a}(\mathbf{q})&=\frac{2}{L^2} \sum_{\mathbf{k}}\sum_{\omega_n} F(i\omega_n, \mathbf{k})F(i\omega_n, \mathbf{k}-\mathbf{q})\nonumber\\
&=\frac{2}{L^2}\!\sum_{\mathbf{k}} 
\frac{\Delta_{\mathbf{k}}}{2E_{\mathbf{k}}}\frac{\Delta_{\mathbf{k}-\mathbf{q}}}{2E_{\mathbf{k}-\mathbf{q}}}
\! \left[\!\frac{1}{E^{-}_{\mathbf{k}-\mathbf{q}}\!+\!E^+_{\mathbf{k}}}
\!+\!\frac{1}{E^{+}_{\mathbf{k}-\mathbf{q}}\!+\!E^{-}_{\mathbf{k}}}\!
\right].
\label{Eq:Pi_a}
\end{align}
For $\Delta_{\mathbf{k}}\equiv 0$, the $\Pi_e(\mathbf{q})$ and $\Pi_h(\mathbf{q})$ reduce to the usual Lindhard functions of the normal state, and $\Pi_a(\mathbf{q})$ vanishes.  

The expression for the static screened electron-hole interaction for unequal masses is,
%
\begin{equation}
V^{sc}_{\mathbf{q}}=
\frac{V^{eh}_\mathbf{q}} 
{1-[\Pi_e(\mathbf{q})V^{ee}_\mathbf{q}+ \Pi_h(\mathbf{q})V^{hh}_\mathbf{q}]
+ 2V^{eh}_\mathbf{q}\Pi_a(\mathbf{q})
+[V^{ee}_\mathbf{q} V^{hh}_\mathbf{q}-(V^{eh}_\mathbf{q})^2]
[\Pi_e(\mathbf{q})\Pi_h(\mathbf{q})-\Pi_a^2(\mathbf{q})]} \ .
\label{Eq:VeffSF}
\end{equation}
%
$V^{ee}_{\mathbf{q}}$ ($V^{hh}_{\mathbf{q}}$) is the bare electron (hole) Coulomb repulsion within one quantum well, and $V^{eh}_{\mathbf{q}}$ is the bare attraction between the electrons and holes in opposite quantum wells:
\begin{align}
V^{ee}_{\mathbf{q}}
=\frac{2\pi e^2}{\epsilon}\frac{1}{|\mathbf{q}|} \mathcal{F}^{ee}_\mathbf{q} \ ; 
\qquad
V^{hh}_{\mathbf{q}}
=\frac{2\pi e^2}{\epsilon}\frac{1}{|\mathbf{q}|} \mathcal{F}^{hh}_\mathbf{q} \ ; 
\qquad 
V^{eh}_{\mathbf{q}}
=-\frac{2\pi e^2}{\epsilon}\frac{e^{-d_c |\mathbf{q}|}}{|\mathbf{q}|} \mathcal{F}^{eh}_\mathbf{q} \ .
\label{Eq:bare_interactions}
\end{align}
We take for the separation parameter $d_c$ the distance between the centre of the two wells.  
The form-factors $\mathcal{F}_\mathbf{q}$ account for the density distribution of the electrons and holes within their respective finite-width wells\cite{Jauho1993}.

We self-consistently solve the superfluid gap equation Eq.\ (\ref{Eq:Delta}), the density equation Eq.\ (\ref{Eq:density}), 
and the screened interaction in the presence of the superfluid Eq.\  (\ref{Eq:VeffSF}) iteratively,
calculating the polarizabilities (Eqs.\ (\ref{Eq:Pi_e}) -- (\ref{Eq:Pi_a})) using the superfluid gaps determined in the preceding iteration. 

\subsection*{Transition temperature}
\vskip-\baselineskip
The superfluid transition temperature in this quasi-2D system is determined as a Berezinskii-Kosterlitz-Thouless (BKT) transition\cite{Kosterlitz1973}.   
For parabolic bands the transition temperature $T_c^{BKT}$ is well approximated by\cite{Botelho2006},
\begin{equation}
T_c^{BKT} = \frac{\pi}{2}\rho_s(T_c^{BKT}) \ .
\label{T_KT}
\end{equation}
Within mean-field theory the superfluid stiffness at zero temperature $\rho_s(0)=\hbar^2n/8 m^\star_r$ depends only on the carrier density $n$, independent of the pair coupling strength.  
We are able to neglect the temperature dependence of $\rho_s(T)$ in Eq.\ (\ref{T_KT}) since $\rho_s(T)$ is approximately constant for temperatures $T\ll \Delta_{\mathrm{max}}$.
Thus the transition temperature is linearly proportional to the carrier density,
\begin{equation}
T_c^{BKT} =  \frac{\hbar^2}{16m^\star_r}\pi n \ .
\label{T_KT_n}
\end{equation}

\section*{Data Availability}
\vskip-\baselineskip
All data generated or analysed during this study are included in this published article and its supplementary information files.

\section*{Acknowledgements}
\vskip-\baselineskip
S.C. acknowledges support of a postdoctoral fellowship from the Flemish Science Foundation (FWO-Vl). 
G.S.acknowledges support from a projectruimte grant associated with the Netherlands Organization of Scientific Research (NWO).
The work was partially supported by the Australian Government through the Australian Research Council Centre of Excellence in Future Low-Energy Electronics (Project No. CE170100039). 

\section*{Author Contributions}
\vskip-\baselineskip
S.C., A.R.H., D.N. and G.S. conceived the idea; 
G.S. designed the Si/Ge material stack and device architecture; 
D.N. and F.M.P. supervised the project; 
S.C., A.P. and S.S.P. developed the theoretical framework, and 
S.C., S.S.P. and M.V. carried out the calculations.  
All authors contributed to the analysis and interpretation of the results, and 
S.C., A.R.H, D.N. and G.S. wrote the paper with input from all authors.

\section*{COMPETING INTERESTS}
\vskip-\baselineskip
Competing financial interests: The authors declare no competing financial interests.

\section*{Supplementary discussion}
\vskip-\baselineskip

\begin{figure*}[h]
\begin{center}
\includegraphics[angle=0,width=0.8\textwidth] {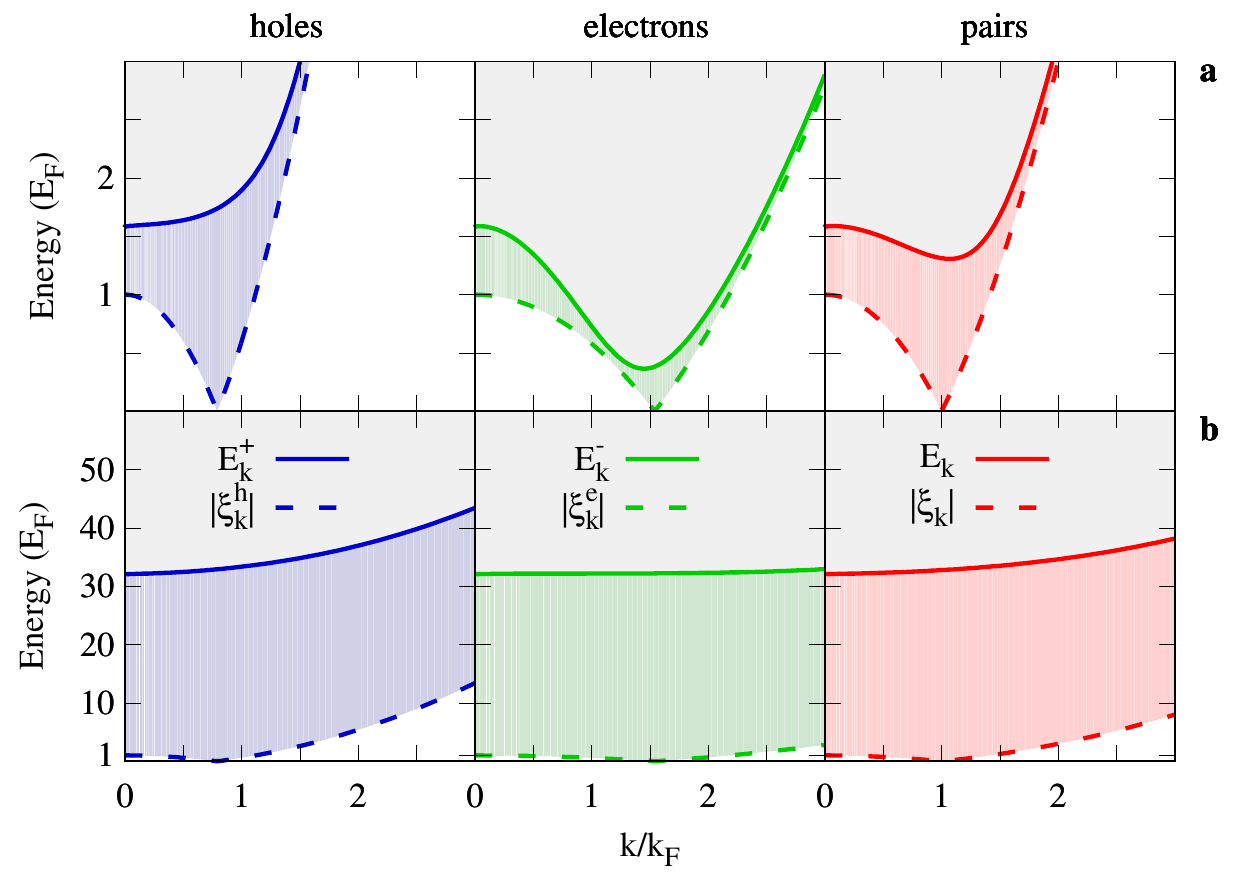}
\end{center}
\caption{\textbf{
Excitation energies for the normal and superfluid states for three 
fixed densities.} 
The normal state electron (hole) single-particle excitations 
$\xi^{e}_k$ ($\xi^{h}_k$), and  corresponding modified 
superfluid state excitations $E^+_k$ ($E^-_k$), scaled to the average 
Fermi energy $E_F$ (with the reduced mass).  
$\xi_k$ and $E_k$ are the corresponding averages.   
Densities:  
\textbf{a}\ \ $n\simeq n_0=5.9 \times 10^{10}$ cm$^{-2}$; 
\textbf{b}\ \ $n=0.5 \times 10^{10}$ cm$^{-2}$.
 }
\label{Fig:energies}
\end{figure*}

We discuss here the origin of the differences in the functional behaviour of the electron and the hole polarizabilities in the superfluid state when the electron and hole masses are different.
The behaviour of $\Pi_e(\mathbf{q})$ and $\Pi_h(\mathbf{q})$ is driven by the changes that the superfluid gap imposes on the excitation spectrum in going from the normal to the superfluid state.  

Supplementary Figure\ \ref{Fig:energies} compares the normal state spectrum $\xi^{e}_k$ ($\xi^{h}_k$) for the electron (hole) single-particle excitations with the corresponding superfluid state excitation spectrum $E^-_k$ ($E^+_k$) (Eqs.\ \ref{Eq:EnergyTerms}).  
The colour-coded shaded areas show the low-energy states in $\xi^{e,h}_k$ that are excluded by the gap from $E^\pm_k$, and thus cannot contribute to the polarizability in the superfluid state. 
It is this exclusion that weakens the screening.
Figures \ref{Fig:energies}a and \ref{Fig:energies}b are for the densities corresponding to Figs.\ \ref{Fig:Pol}a and \ref{Fig:Pol}b in the main text. 

We recall in Fig.\ \ref{Fig:Pol}a of the main text  near the onset density, that $\Pi_e(\mathbf{q})$ initially grows with increasing $\mathbf{q}$, passes through a maximum, and then goes slowly to zero, while $\Pi_h(\mathbf{q})$ decreases monotonically to zero.
We can deduce the cause of this strikingly different behaviour from Fig.\ \ref{Fig:energies}a.  
The right panel for the pairs shows the familiar behaviour of $E_k= \sqrt{\xi_k^2 + \Delta_k^{2}}$, where $\xi_{\mathbf{k}}=\frac{1}{2}\left(\xi^e_{\mathbf{k}} + \xi^h_{\mathbf{k}}\right)$,  
which goes through a minimum at $k=k_F$ at an energy equal to $\Delta_k$.  
In our system, the $E_k$ is modified by the positive $\delta \xi_k=\frac{1}{2}\left(\xi^h_{\mathbf{k}} - \xi^e_{\mathbf{k}}\right)$, which takes into account the unequal masses.
The middle panel for electrons shows $E^-_k= E_k - \delta \xi_k$ passing through a minimum that is lower than for $E_k$, because of the presence of $\delta \xi_k$.  
The pronounced  minimum in $E^-_k$  leads to the maximum seen in $\Pi_e(\mathbf{q})$ (Fig.\ 3a of the main text).  
The left panel for holes shows, by contrast, that due to the addition of $\delta \xi_k$, $E^+_k$ has no minimum at all, leading to a $\Pi_h(\mathbf{q})$ that depends monotonically on $\mathbf{q}$. 
The net result is that $E^+_k$ is shifted up relative to the normal state $\xi^{h}_k$, while $E^-_k$ closely tracks the normal state $\xi^e_k$.
This means that the superfluid gap $\Delta_{\mathbf{k}}$ is markedly less effective at blocking the excitation states for electrons than for holes in the range important for screening, $k<2k_F$. 

Figure \ref{Fig:Pol}b of the main text shows that the behaviour of $\Pi_e(\mathbf{q})$ markedly changes when the density is decreased.  
This is because by the time one arrives in the BEC regime, the $\Pi_e(\mathbf{q})$ for the superfluid state has become very similar in behaviour to $\Pi_h(\mathbf{q})$ out to large $\mathbf{q} \lesssim 4k_F$.
Figure\ \ref{Fig:energies}b shows in the BEC regime that the very large $\Delta_k$, on the scale of $E_F$, excludes a huge number of low-lying excited states from participating in the screening, with $E^\pm_k$ shifted up in energy relative to $\xi^{e,h}_k$ by more than $30E_F$. 
Since here $\Delta_k\gg \delta \xi_k$, it follows that $E_k\simeq E^-_k\simeq E^+_k$, so the unequal masses no longer differentiate the polarizability properties.

\bigskip

\begin{thebibliography}{56}
\makeatletter
\providecommand \@ifxundefined [1]{%
 \@ifx{#1\undefined}
}%
\providecommand \@ifnum [1]{%
 \ifnum #1\expandafter \@firstoftwo
 \else \expandafter \@secondoftwo
 \fi
}%
\providecommand \@ifx [1]{%
 \ifx #1\expandafter \@firstoftwo
 \else \expandafter \@secondoftwo
 \fi
}%
\providecommand \natexlab [1]{#1}%
\providecommand \enquote  [1]{``#1''}%
\providecommand \bibnamefont  [1]{#1}%
\providecommand \bibfnamefont [1]{#1}%
\providecommand \citenamefont [1]{#1}%
\providecommand \href@noop [0]{\@secondoftwo}%
\providecommand \href [0]{\begingroup \@sanitize@url \@href}%
\providecommand \@href[1]{\@@startlink{#1}\@@href}%
\providecommand \@@href[1]{\endgroup#1\@@endlink}%
\providecommand \@sanitize@url [0]{\catcode `\\12\catcode `\$12\catcode
  `\&12\catcode `\#12\catcode `\^12\catcode `\_12\catcode `\%12\relax}%
\providecommand \@@startlink[1]{}%
\providecommand \@@endlink[0]{}%
\providecommand \url  [0]{\begingroup\@sanitize@url \@url }%
\providecommand \@url [1]{\endgroup\@href {#1}{\urlprefix }}%
\providecommand \urlprefix  [0]{URL }%
\providecommand \Eprint [0]{\href }%
\providecommand \doibase [0]{http://dx.doi.org/}%
\providecommand \selectlanguage [0]{\@gobble}%
\providecommand \bibinfo  [0]{\@secondoftwo}%
\providecommand \bibfield  [0]{\@secondoftwo}%
\providecommand \translation [1]{[#1]}%
\providecommand \BibitemOpen [0]{}%
\providecommand \bibitemStop [0]{}%
\providecommand \bibitemNoStop [0]{.\EOS\space}%
\providecommand \EOS [0]{\spacefactor3000\relax}%
\providecommand \BibitemShut  [1]{\csname bibitem#1\endcsname}%
\let\auto@bib@innerbib\@empty
\bibitem{Lozovik1975}
\bibinfo{author}{Lozovik, Y.~E.} \& \bibinfo{author}{Yudson, V.~I.}
\newblock \bibinfo{title}{Feasibility of superfluidity of paired spatially
  separated electrons and holes}.
\newblock \emph{\bibinfo{journal}{JETP Lett. (USSR)}}
  \textbf{\bibinfo{volume}{22}}, \bibinfo{pages}{274} (\bibinfo{year}{1975}).
\newblock \urlprefix\url{http://jetpletters.ac.ru/ps/1530/article_23399.pdf}.
\newblock \bibinfo{note}{(Pis’ma Zh.\ Eksp.\ Teor.\ Fiz.\ {\bf 22}, 556
  (1975))}.

\bibitem{Lozovik1976}
\bibinfo{author}{Lozovik, Y.~E.} \& \bibinfo{author}{Yudson, V.~I.}
\newblock \bibinfo{title}{A new mechanism for superconductivity: pairing
  between spatially separated electrons and holes}.
\newblock \emph{\bibinfo{journal}{Sov. Phys. JETP}}
  \textbf{\bibinfo{volume}{44}}, \bibinfo{pages}{389} (\bibinfo{year}{1976}).
\newblock \urlprefix\url{http://www.jetp.ac.ru/cgi-bin/dn/e_044_02_0389.pdf}.
\newblock \bibinfo{note}{(Zh. Eksp. Teor. Fiz. {\bf 71}, 738 (1976))}.

\bibitem{Su2008}
\bibinfo{author}{Su, J.-J.} \& \bibinfo{author}{MacDonald, A.~H.}
\newblock \bibinfo{title}{How to make a bilayer exciton condensate flow}.
\newblock \emph{\bibinfo{journal}{Nat. Phys.}} \textbf{\bibinfo{volume}{4}},
  \bibinfo{pages}{799} (\bibinfo{year}{2008}).
\newblock \urlprefix\url{https://www.nature.com/articles/nphys1055}.

\bibitem{Nandi2012}
\bibinfo{author}{Nandi, D.}, \bibinfo{author}{Finck, A. D.~K.},
  \bibinfo{author}{Eisenstein, J.~P.}, \bibinfo{author}{Pfeiffer, L.~N.} \&
  \bibinfo{author}{West, K.~W.}
\newblock \bibinfo{title}{Exciton condensation and perfect {C}oulomb drag}.
\newblock \emph{\bibinfo{journal}{Nature}} \textbf{\bibinfo{volume}{488}},
  \bibinfo{pages}{481} (\bibinfo{year}{2012}).
\newblock \urlprefix\url{https://www.nature.com/articles/nature11302}.

\bibitem{Pieri2007}
\bibinfo{author}{Pieri, P.}, \bibinfo{author}{Neilson, D.} \&
  \bibinfo{author}{Strinati, G.~C.}
\newblock \bibinfo{title}{Effects of density imbalance on the {BCS-BEC}
  crossover in semiconductor electron-hole bilayers}.
\newblock \emph{\bibinfo{journal}{Phys. Rev. B}} \textbf{\bibinfo{volume}{75}},
  \bibinfo{pages}{113301} (\bibinfo{year}{2007}).
\newblock \urlprefix\url{https://link.aps.org/doi/10.1103/PhysRevB.75.113301}.

\bibitem{Wang2017a}
\bibinfo{author}{Wang, J.}, \bibinfo{author}{Che, Y.}, \bibinfo{author}{Zhang,
  L.} \& \bibinfo{author}{Chen, Q.}
\newblock \bibinfo{title}{Enhancement effect of mass imbalance on
  {F}ulde-{F}errell-{L}arkin-{O}vchinnikov type of pairing in {F}ermi-{F}ermi
  mixtures of ultracold quantum gases}.
\newblock \emph{\bibinfo{journal}{Sci. Rep.}} \textbf{\bibinfo{volume}{7}},
  \bibinfo{pages}{39783} (\bibinfo{year}{2017}).
\newblock \urlprefix\url{https://www.nature.com/articles/srep39783}.

\bibitem{Forbes2005}
\bibinfo{author}{Forbes, M.~M.}, \bibinfo{author}{Gubankova, E.},
  \bibinfo{author}{Liu, W.~V.} \& \bibinfo{author}{Wilczek, F.}
\newblock \bibinfo{title}{Stability criteria for breached-pair superfluidity}.
\newblock \emph{\bibinfo{journal}{Phys. Rev. Lett.}}
  \textbf{\bibinfo{volume}{94}}, \bibinfo{pages}{017001}
  (\bibinfo{year}{2005}).
\newblock
  \urlprefix\url{https://journals.aps.org/prl/abstract/10.1103/PhysRevLett.94.017001}.

\bibitem{Kinnunen2018}
\bibinfo{author}{Kinnunen, J.~J.}, \bibinfo{author}{Baarsma, J.~E.},
  \bibinfo{author}{Martikainen, J.-P.} \& \bibinfo{author}{Törmä, P.}
\newblock \bibinfo{title}{The {F}ulde-{F}errell-{L}arkin-{O}vchinnikov state
  for ultracold fermions in lattice and harmonic potentials: A review}.
\newblock \emph{\bibinfo{journal}{Rep. Prog. Phys.}}
  \textbf{\bibinfo{volume}{81}}, \bibinfo{pages}{046401}
  (\bibinfo{year}{2018}).
\newblock \urlprefix\url{https://doi.org/10.1088%2F1361-6633%2Faaa4ad}.

\bibitem{Frank2018}
\bibinfo{author}{Frank, B.}, \bibinfo{author}{Lang, J.} \&
  \bibinfo{author}{Zwerger, W.}
\newblock \bibinfo{title}{Universal phase diagram and scaling functions of
  imbalanced {F}ermi gases}.
\newblock \emph{\bibinfo{journal}{J. Exp. Theor. Phys.}}
  \textbf{\bibinfo{volume}{127}}, \bibinfo{pages}{812} (\bibinfo{year}{2018}).
\newblock \urlprefix\url{https://doi.org/10.1134/S1063776118110031}.

\bibitem{Perali2013}
\bibinfo{author}{Perali, A.}, \bibinfo{author}{Neilson, D.} \&
  \bibinfo{author}{Hamilton, A.~R.}
\newblock \bibinfo{title}{High-temperature superfluidity in double-bilayer
  graphene}.
\newblock \emph{\bibinfo{journal}{Phys. Rev. Lett.}}
  \textbf{\bibinfo{volume}{110}}, \bibinfo{pages}{146803}
  (\bibinfo{year}{2013}).
\newblock
  \urlprefix\url{https://link.aps.org/doi/10.1103/PhysRevLett.110.146803}.

\bibitem{Fogler2014}
\bibinfo{author}{Fogler, M.~M.}, \bibinfo{author}{Butov, L.~V.} \&
  \bibinfo{author}{Novoselov, K.~S.}
\newblock \bibinfo{title}{High-temperature superfluidity with indirect excitons
  in van der {W}aals heterostructures}.
\newblock \emph{\bibinfo{journal}{Nat. Commun.}} \textbf{\bibinfo{volume}{5}},
  \bibinfo{pages}{4555} (\bibinfo{year}{2014}).
\newblock \urlprefix\url{https://www.nature.com/articles/ncomms5555}.

\bibitem{Burg2018}
\bibinfo{author}{Burg, G.~W.} \emph{et~al.}
\newblock \bibinfo{title}{Strongly enhanced tunneling at total charge
  neutrality in double-bilayer graphene-{${\mathrm{WSe}}_{2}$}
  heterostructures}.
\newblock \emph{\bibinfo{journal}{Phys. Rev. Lett.}}
  \textbf{\bibinfo{volume}{120}}, \bibinfo{pages}{177702}
  (\bibinfo{year}{2018}).
\newblock
  \urlprefix\url{https://link.aps.org/doi/10.1103/PhysRevLett.120.177702}.

\bibitem{Wang2019}
\bibinfo{author}{Wang, Z.} \emph{et~al.}
\newblock \bibinfo{title}{Evidence of high-temperature exciton condensation in
  two-dimensional atomic double layers}.
\newblock \emph{\bibinfo{journal}{Nature (London)}}
  \textbf{\bibinfo{volume}{574}}, \bibinfo{pages}{76} (\bibinfo{year}{2019}).
\newblock \urlprefix\url{https://doi.org/10.1038/s41586-019-1591-7}.

\bibitem{Frisenda2018}
\bibinfo{author}{Frisenda, R.} \emph{et~al.}
\newblock \bibinfo{title}{Recent progress in the assembly of nanodevices and
  van der {W}aals heterostructures by deterministic placement of {2D}
  materials}.
\newblock \emph{\bibinfo{journal}{Chem. Soc. Rev.}}
  \textbf{\bibinfo{volume}{47}}, \bibinfo{pages}{53} (\bibinfo{year}{2018}).
\newblock \urlprefix\url{http://dx.doi.org/10.1039/C7CS00556C}.

\bibitem{Croxall2008}
\bibinfo{author}{Croxall, A.~F.} \emph{et~al.}
\newblock \bibinfo{title}{Anomalous {C}oulomb drag in electron-hole bilayers}.
\newblock \emph{\bibinfo{journal}{Phys. Rev. Lett.}}
  \textbf{\bibinfo{volume}{101}}, \bibinfo{pages}{246801}
  (\bibinfo{year}{2008}).
\newblock
  \urlprefix\url{https://link.aps.org/doi/10.1103/PhysRevLett.101.246801}.

\bibitem{Seamons2009}
\bibinfo{author}{Seamons, J.~A.}, \bibinfo{author}{Morath, C.~P.},
  \bibinfo{author}{Reno, J.~L.} \& \bibinfo{author}{Lilly, M.~P.}
\newblock \bibinfo{title}{Coulomb drag in the exciton regime in electron-hole
  bilayers}.
\newblock \emph{\bibinfo{journal}{Phys. Rev. Lett.}}
  \textbf{\bibinfo{volume}{102}}, \bibinfo{pages}{026804}
  (\bibinfo{year}{2009}).
\newblock
  \urlprefix\url{https://link.aps.org/doi/10.1103/PhysRevLett.102.026804}.

\bibitem{SaberiPouya2020}
\bibinfo{author}{Saberi-Pouya, S.} \emph{et~al.}
\newblock \bibinfo{title}{Experimental conditions for the observation of
  electron-hole superfluidity in {GaAs} heterostructures}.
\newblock \emph{\bibinfo{journal}{Phys. Rev. B}}
  \textbf{\bibinfo{volume}{101}}, \bibinfo{pages}{140501(R)}
  (\bibinfo{year}{2020}).
\newblock \urlprefix\url{https://link.aps.org/doi/10.1103/PhysRevB.101.140501}.

\bibitem{DasGupta2011a}
\bibinfo{author}{Das~Gupta, K.} \emph{et~al.}
\newblock \bibinfo{title}{Experimental progress towards probing the ground
  state of an electron-hole bilayer by low-temperature transport}.
\newblock \emph{\bibinfo{journal}{Advances in Condensed Matter Physics}}
  \textbf{\bibinfo{volume}{2011}} (\bibinfo{year}{2011}).
\newblock \urlprefix\url{https://www.hindawi.com/journals/acmp/2011/727958/}.

\bibitem{schaffler_high-mobility_1997}
\bibinfo{author}{Schäffler, F.}
\newblock \bibinfo{title}{High-mobility {Si} and {Ge} structures}.
\newblock \emph{\bibinfo{journal}{Semicond. Sci. Technol.}}
  \textbf{\bibinfo{volume}{12}}, \bibinfo{pages}{1515} (\bibinfo{year}{1997}).
\newblock
  \urlprefix\url{http://stacks.iop.org/0268-1242/12/i=12/a=001?key=crossref.83bc7e505a3f624490fde29ed7c292e4}.

\bibitem{lee_strained_2004}
\bibinfo{author}{Lee, M.~L.}, \bibinfo{author}{Fitzgerald, E.~A.},
  \bibinfo{author}{Bulsara, M.~T.}, \bibinfo{author}{Currie, M.~T.} \&
  \bibinfo{author}{Lochtefeld, A.}
\newblock \bibinfo{title}{Strained {Si}, {SiGe}, and {Ge} channels for
  high-mobility metal-oxide-semiconductor field-effect transistors}.
\newblock \emph{\bibinfo{journal}{J. Appl. Phys.}}
  \textbf{\bibinfo{volume}{97}}, \bibinfo{pages}{011101}
  (\bibinfo{year}{2004}).
\newblock \urlprefix\url{https://aip.scitation.org/doi/10.1063/1.1819976}.

\bibitem{virgilio_type-i_2006}
\bibinfo{author}{Virgilio, M.} \& \bibinfo{author}{Grosso, G.}
\newblock \bibinfo{title}{Type-{I} alignment and direct fundamental gap in
  {SiGe} based heterostructures}.
\newblock \emph{\bibinfo{journal}{J. Phys. Condens. Mat.}}
  \textbf{\bibinfo{volume}{18}}, \bibinfo{pages}{1021} (\bibinfo{year}{2006}).
\newblock
  \urlprefix\url{https://iopscience.iop.org/article/10.1088/0953-8984/18/3/018}.

\bibitem{lu_observation_2009}
\bibinfo{author}{Lu, T.~M.}, \bibinfo{author}{Tsui, D.~C.},
  \bibinfo{author}{Lee, C.-H.} \& \bibinfo{author}{Liu, C.~W.}
\newblock \bibinfo{title}{Observation of two-dimensional electron gas in a {Si}
  quantum well with mobility of 1.6×10$^6$~cm$^2$/{Vs}}.
\newblock \emph{\bibinfo{journal}{Appl. Phys. Lett.}}
  \textbf{\bibinfo{volume}{94}}, \bibinfo{pages}{182102}
  (\bibinfo{year}{2009}).
\newblock \urlprefix\url{https://aip.scitation.org/doi/10.1063/1.3127516}.

\bibitem{dobbie_ultra-high_2012}
\bibinfo{author}{Dobbie, A.} \emph{et~al.}
\newblock \bibinfo{title}{Ultra-high hole mobility exceeding one million in a
  strained germanium quantum well}.
\newblock \emph{\bibinfo{journal}{Appl. Phys. Lett.}}
  \textbf{\bibinfo{volume}{101}}, \bibinfo{pages}{172108}
  (\bibinfo{year}{2012}).
\newblock \urlprefix\url{https://aip.scitation.org/doi/10.1063/1.4763476}.

\bibitem{wuetz_multiplexed_2020}
\bibinfo{author}{Paquelet~Wuetz, B.} \emph{et~al.}
\newblock \bibinfo{title}{Multiplexed quantum transport using commercial
  off-the-shelf {CMOS} at sub-kelvin temperatures}.
\newblock \emph{\bibinfo{journal}{npj Quantum Inf.}}
  \textbf{\bibinfo{volume}{6}}, \bibinfo{pages}{43} (\bibinfo{year}{2020}).
\newblock \urlprefix\url{https://www.nature.com/articles/s41534-020-0274-4}.

\bibitem{lodari_low_2020}
\bibinfo{author}{Lodari, M.} \emph{et~al.}
\newblock \bibinfo{title}{Low percolation density and charge noise with holes
  in germanium}  (\bibinfo{year}{2020}).
\newblock \urlprefix\url{http://arxiv.org/abs/2007.06328}.
\newblock \bibinfo{note}{Preprint at: http://arxiv.org/abs/2007.06328}.

\bibitem{Vandersypen2017InterfacingCoherent}
\bibinfo{author}{Vandersypen, L. M.~K.} \emph{et~al.}
\newblock \bibinfo{title}{{Interfacing spin qubits in quantum dots and
  donors-hot, dense, and coherent}}.
\newblock \emph{\bibinfo{journal}{npj Quantum Inf.}}
  \textbf{\bibinfo{volume}{3}}, \bibinfo{pages}{34} (\bibinfo{year}{2017}).
\newblock \urlprefix\url{https://www.nature.com/articles/s41534-017-0038-y}.

\bibitem{scappucci_germanium_2020}
\bibinfo{author}{Scappucci, G.} \emph{et~al.}
\newblock \bibinfo{title}{The germanium quantum information route}
  (\bibinfo{year}{2020}).
\newblock \urlprefix\url{http://arxiv.org/abs/2004.08133}.
\newblock \bibinfo{note}{Preprint at: http://arxiv.org/abs/2004.08133}.

\bibitem{Yoneda2018A99.9}
\bibinfo{author}{Yoneda, J.} \emph{et~al.}
\newblock \bibinfo{title}{{A quantum-dot spin qubit with coherence limited by
  charge noise and fidelity higher than 99.9{\%}}}.
\newblock \emph{\bibinfo{journal}{Nature Nanotech}}
  \textbf{\bibinfo{volume}{13}}, \bibinfo{pages}{102} (\bibinfo{year}{2018}).
\newblock \urlprefix\url{http://dx.doi.org/10.1038/s41565-017-0014-x}.

\bibitem{watson_programmable_2018}
\bibinfo{author}{Watson, T.~F.} \emph{et~al.}
\newblock \bibinfo{title}{A programmable two-qubit quantum processor in
  silicon}.
\newblock \emph{\bibinfo{journal}{Nature}} \textbf{\bibinfo{volume}{555}},
  \bibinfo{pages}{633} (\bibinfo{year}{2018}).
\newblock \urlprefix\url{http://www.nature.com/articles/nature25766}.

\bibitem{hendrickx_four-qubit_2020}
\bibinfo{author}{Hendrickx, N.~W.} \emph{et~al.}
\newblock \bibinfo{title}{A four-qubit germanium quantum processor}
  (\bibinfo{year}{2020}).
\newblock \urlprefix\url{http://arxiv.org/abs/2009.04268}.
\newblock \bibinfo{note}{Preprint at: http://arxiv.org/abs/2009.04268}.

\bibitem{hendrickx_fast_2020}
\bibinfo{author}{Hendrickx, N.~W.}, \bibinfo{author}{Franke, D.~P.},
  \bibinfo{author}{Sammak, A.}, \bibinfo{author}{Scappucci, G.} \&
  \bibinfo{author}{Veldhorst, M.}
\newblock \bibinfo{title}{Fast two-qubit logic with holes in germanium}.
\newblock \emph{\bibinfo{journal}{Nature}} \textbf{\bibinfo{volume}{577}},
  \bibinfo{pages}{487} (\bibinfo{year}{2020}).
\newblock \urlprefix\url{http://www.nature.com/articles/s41586-019-1919-3}.

\bibitem{hendrickx_gate-controlled_2018}
\bibinfo{author}{Hendrickx, N.~W.} \emph{et~al.}
\newblock \bibinfo{title}{Gate-controlled quantum dots and superconductivity in
  planar germanium}.
\newblock \emph{\bibinfo{journal}{Nature Commun.}}
  \textbf{\bibinfo{volume}{9}}, \bibinfo{pages}{1} (\bibinfo{year}{2018}).
\newblock \urlprefix\url{https://www.nature.com/articles/s41467-018-05299-x/}.

\bibitem{hendrickx_ballistic_2019}
\bibinfo{author}{Hendrickx, N.~W.} \emph{et~al.}
\newblock \bibinfo{title}{Ballistic supercurrent discretization and
  micrometer-long {Josephson} coupling in germanium}.
\newblock \emph{\bibinfo{journal}{Phys. Rev. B}} \textbf{\bibinfo{volume}{99}}
  (\bibinfo{year}{2019}).
\newblock \urlprefix\url{https://doi.org/10.1103%2Fphysrevb.99.075435}.

\bibitem{vigneau_germanium_2019}
\bibinfo{author}{Vigneau, F.} \emph{et~al.}
\newblock \bibinfo{title}{Germanium {Quantum}-{Well} {Josephson}
  {Field}-{Effect} {Transistors} and {Interferometers}}.
\newblock \emph{\bibinfo{journal}{Nano Lett.}} \textbf{\bibinfo{volume}{19}},
  \bibinfo{pages}{1023} (\bibinfo{year}{2019}).
\newblock \urlprefix\url{https://doi.org/10.1021%2Facs.nanolett.8b04275}.

\bibitem{lodari_light_2019}
\bibinfo{author}{Lodari, M.} \emph{et~al.}
\newblock \bibinfo{title}{Light effective hole mass in undoped {Ge/SiGe}
  quantum wells}.
\newblock \emph{\bibinfo{journal}{Phys. Rev. B}}
  \textbf{\bibinfo{volume}{100}}, \bibinfo{pages}{041304}
  (\bibinfo{year}{2019}).
\newblock \urlprefix\url{https://link.aps.org/doi/10.1103/PhysRevB.100.041304}.

\bibitem{Zwanenburg2013}
\bibinfo{author}{Zwanenburg, F.~A.} \emph{et~al.}
\newblock \bibinfo{title}{Silicon quantum electronics}.
\newblock \emph{\bibinfo{journal}{Rev. Mod. Phys.}}
  \textbf{\bibinfo{volume}{85}}, \bibinfo{pages}{961} (\bibinfo{year}{2013}).
\newblock \urlprefix\url{https://link.aps.org/doi/10.1103/RevModPhys.85.961}.

\bibitem{Pillarisetty2011Academic}
\bibinfo{author}{Pillarisetty, R.}
\newblock \bibinfo{title}{Academic and industry research progress in germanium
  nanodevices}.
\newblock \emph{\bibinfo{journal}{Nature}} \textbf{\bibinfo{volume}{479}},
  \bibinfo{pages}{324} (\bibinfo{year}{2011}).
\newblock \urlprefix\url{https://www.nature.com/articles/nature10678}.

\bibitem{matthews_defects_1976}
\bibinfo{author}{Matthews, J.~W.} \& \bibinfo{author}{Blakeslee, A.~E.}
\newblock \bibinfo{title}{Defects in epitaxial multilayers: {III}.
  {Preparation} of almost perfect multilayers}.
\newblock \emph{\bibinfo{journal}{J. Cryst. Growth}}
  \textbf{\bibinfo{volume}{32}}, \bibinfo{pages}{265} (\bibinfo{year}{1976}).
\newblock
  \urlprefix\url{http://www.sciencedirect.com/science/article/pii/0022024876900415}.

\bibitem{Paul2010}
\bibinfo{author}{Paul, D.}
\newblock \bibinfo{title}{The progress towards terahertz quantum cascade lasers
  on silicon substrates}.
\newblock \emph{\bibinfo{journal}{Laser \& Photonics Rev.}}
  \textbf{\bibinfo{volume}{4}}, \bibinfo{pages}{610} (\bibinfo{year}{2010}).
\newblock
  \urlprefix\url{https://onlinelibrary.wiley.com/doi/abs/10.1002/lpor.200910038}.

\bibitem{sammak_shallow_2019}
\bibinfo{author}{Sammak, A.} \emph{et~al.}
\newblock \bibinfo{title}{Shallow and {Undoped} {Germanium} {Quantum} {Wells}:
  {A} {Playground} for {Spin} and {Hybrid} {Quantum} {Technology}}.
\newblock \emph{\bibinfo{journal}{Adv. Funct. Mater.}} \bibinfo{pages}{1807613}
  (\bibinfo{year}{2019}).
\newblock \urlprefix\url{https://doi.org/10.1002%2Fadfm.201807613}.

\bibitem{dyck_accurate_2017}
\bibinfo{author}{Dyck, O.} \emph{et~al.}
\newblock \bibinfo{title}{Accurate {Quantification} of {Si}/{SiGe} {Interface}
  {Profiles} via {Atom} {Probe} {Tomography}}.
\newblock \emph{\bibinfo{journal}{Adv. Mater. Interfaces}}
  \textbf{\bibinfo{volume}{4}}, \bibinfo{pages}{1700622}
  (\bibinfo{year}{2017}).
\newblock
  \urlprefix\url{https://onlinelibrary.wiley.com/doi/abs/10.1002/admi.201700622}.

\bibitem{shah_reverse_2008}
\bibinfo{author}{Shah, V.~A.} \emph{et~al.}
\newblock \bibinfo{title}{Reverse graded relaxed buffers for high {Ge} content
  {SiGe} virtual substrates}.
\newblock \emph{\bibinfo{journal}{Appl. Phys. Lett.}}
  \textbf{\bibinfo{volume}{93}}, \bibinfo{pages}{192103}
  (\bibinfo{year}{2008}).
\newblock \urlprefix\url{https://aip.scitation.org/doi/10.1063/1.3023068}.

\bibitem{su_how_2008}
\bibinfo{author}{Su, J.-J.} \& \bibinfo{author}{MacDonald, A.~H.}
\newblock \bibinfo{title}{How to make a bilayer exciton condensate flow}.
\newblock \emph{\bibinfo{journal}{Nature Physics}}
  \textbf{\bibinfo{volume}{4}}, \bibinfo{pages}{799} (\bibinfo{year}{2008}).
\newblock \urlprefix\url{https://www.nature.com/articles/nphys1055}.

\bibitem{Lozovik2012}
\bibinfo{author}{Lozovik, Y.~E.}, \bibinfo{author}{Ogarkov, S.~L.} \&
  \bibinfo{author}{Sokolik, A.~A.}
\newblock \bibinfo{title}{Condensation of electron-hole pairs in a two-layer
  graphene system: {C}orrelation effects}.
\newblock \emph{\bibinfo{journal}{Phys. Rev. B}} \textbf{\bibinfo{volume}{86}},
  \bibinfo{pages}{045429} (\bibinfo{year}{2012}).
\newblock \urlprefix\url{https://link.aps.org/doi/10.1103/PhysRevB.86.045429}.

\bibitem{Neilson2014}
\bibinfo{author}{Neilson, D.}, \bibinfo{author}{Perali, A.} \&
  \bibinfo{author}{Hamilton, A.~R.}
\newblock \bibinfo{title}{Excitonic superfluidity and screening in
  electron-hole bilayer systems}.
\newblock \emph{\bibinfo{journal}{Phys. Rev. B}} \textbf{\bibinfo{volume}{89}},
  \bibinfo{pages}{060502(R)} (\bibinfo{year}{2014}).
\newblock \urlprefix\url{https://link.aps.org/doi/10.1103/PhysRevB.89.060502}.

\bibitem{LopezRios2018}
\bibinfo{author}{L\'opez~R\'{\i}os, P.}, \bibinfo{author}{Perali, A.},
  \bibinfo{author}{Needs, R.~J.} \& \bibinfo{author}{Neilson, D.}
\newblock \bibinfo{title}{Evidence from quantum {M}onte {C}arlo simulations of
  large-gap superfluidity and {BCS}-{BEC} crossover in double electron-hole
  layers}.
\newblock \emph{\bibinfo{journal}{Phys. Rev. Lett.}}
  \textbf{\bibinfo{volume}{120}}, \bibinfo{pages}{177701}
  (\bibinfo{year}{2018}).
\newblock
  \urlprefix\url{https://link.aps.org/doi/10.1103/PhysRevLett.120.177701}.

\bibitem{Salasnich2005}
\bibinfo{author}{Salasnich, L.}, \bibinfo{author}{Manini, N.} \&
  \bibinfo{author}{Parola, A.}
\newblock \bibinfo{title}{Condensate fraction of a {F}ermi gas in the {BCS-BEC}
  crossover}.
\newblock \emph{\bibinfo{journal}{Phys. Rev. A}} \textbf{\bibinfo{volume}{72}},
  \bibinfo{pages}{023621} (\bibinfo{year}{2005}).
\newblock
  \urlprefix\url{https://journals.aps.org/pra/abstract/10.1103/PhysRevA.72.023621}.

\bibitem{Guidini2014}
\bibinfo{author}{Guidini, A.} \& \bibinfo{author}{Perali, A.}
\newblock \bibinfo{title}{Band-edge {BCS}{\textendash}{BEC} crossover in a
  two-band superconductor: physical properties and detection parameters}.
\newblock \emph{\bibinfo{journal}{Supercond. Sci. Tech.}}
  \textbf{\bibinfo{volume}{27}}, \bibinfo{pages}{124002}
  (\bibinfo{year}{2014}).
\newblock
  \urlprefix\url{https://iopscience.iop.org/article/10.1088/0953-2048/27/12/124002/meta}.

\bibitem{Kosterlitz1973}
\bibinfo{author}{Kosterlitz, J.~M.} \& \bibinfo{author}{Thouless, D.~J.}
\newblock \bibinfo{title}{Ordering, metastability and phase transitions in
  two-dimensional systems}.
\newblock \emph{\bibinfo{journal}{J. Phys. C: Solid State}}
  \textbf{\bibinfo{volume}{6}}, \bibinfo{pages}{1181} (\bibinfo{year}{1973}).
\newblock \urlprefix\url{https://doi.org/10.1088/0022-3719/6/7/010}.

\bibitem{Butov2004}
\bibinfo{author}{Butov, L.~V.}
\newblock \bibinfo{title}{Condensation and pattern formation in cold exciton
  gases in coupled quantum wells}.
\newblock \emph{\bibinfo{journal}{J. Phys. Condens. Mat.}}
  \textbf{\bibinfo{volume}{16}}, \bibinfo{pages}{R1577} (\bibinfo{year}{2004}).
\newblock \urlprefix\url{https://doi.org/10.1088%2F0953-8984%2F16%2F50%2Fr02}.

\bibitem{VanderDonck2020}
\bibinfo{author}{Van~der Donck, M.} \emph{et~al.}
\newblock \bibinfo{title}{Three-dimensional electron-hole superfluidity in a
  superlattice close to room temperature}.
\newblock \emph{\bibinfo{journal}{Phys. Rev. B}}
  \textbf{\bibinfo{volume}{102}}, \bibinfo{pages}{060503}
  (\bibinfo{year}{2020}).
\newblock \urlprefix\url{https://link.aps.org/doi/10.1103/PhysRevB.102.060503}.

\bibitem{grange_atomic-scale_2020}
\bibinfo{author}{Grange, T.} \emph{et~al.}
\newblock \bibinfo{title}{Atomic-{Scale} {Insights} into {Semiconductor}
  {Heterostructures}: {From} {Experimental} {Three}-{Dimensional} {Analysis} of
  the {Interface} to a {Generalized} {Theory} of {Interfacial} {Roughness}
  {Scattering}}.
\newblock \emph{\bibinfo{journal}{Phys. Rev. Appl.}}
  \textbf{\bibinfo{volume}{13}}, \bibinfo{pages}{044062}
  (\bibinfo{year}{2020}).
\newblock
  \urlprefix\url{https://link.aps.org/doi/10.1103/PhysRevApplied.13.044062}.

\bibitem{reddy_bilayer_2009}
\bibinfo{author}{Reddy, D.}, \bibinfo{author}{Register, L.~F.},
  \bibinfo{author}{Tutuc, E.}, \bibinfo{author}{MacDonald, A.} \&
  \bibinfo{author}{Banerjee, S.~K.}
\newblock \bibinfo{title}{Bilayer {pseudospin} field effect transistor
  ({BiSFET}): {A} proposed logic device and circuits}.
\newblock In \emph{\bibinfo{booktitle}{2009 {Device} {Research} {Conference}}},
  \bibinfo{pages}{67} (\bibinfo{year}{2009}).
\newblock \bibinfo{note}{ISSN: 1548-3770}.

\bibitem{xue_cmos-based_2020}
\bibinfo{author}{Xue, X.} \emph{et~al.}
\newblock \bibinfo{title}{{CMOS}-based cryogenic control of silicon quantum
  circuits}  (\bibinfo{year}{2020}).
\newblock \urlprefix\url{http://arxiv.org/abs/2009.14185}.
\newblock \bibinfo{note}{Preprint at: http://arxiv.org/abs/2009.14185}.

\bibitem{Jauho1993}
\bibinfo{author}{Jauho, A.-P.} \& \bibinfo{author}{Smith, H.}
\newblock \bibinfo{title}{Coulomb drag between parallel two-dimensional
  electron systems}.
\newblock \emph{\bibinfo{journal}{Phys. Rev. B}} \textbf{\bibinfo{volume}{47}},
  \bibinfo{pages}{4420} (\bibinfo{year}{1993}).
\newblock \urlprefix\url{https://link.aps.org/doi/10.1103/PhysRevB.47.4420}.

\bibitem{Botelho2006}
\bibinfo{author}{Botelho, S.~S.} \& \bibinfo{author}{S\'a~de Melo, C. A.~R.}
\newblock \bibinfo{title}{Vortex-antivortex lattice in ultracold fermionic
  gases}.
\newblock \emph{\bibinfo{journal}{Phys. Rev. Lett.}}
  \textbf{\bibinfo{volume}{96}}, \bibinfo{pages}{040404}
  (\bibinfo{year}{2006}).
\newblock
  \urlprefix\url{https://link.aps.org/doi/10.1103/PhysRevLett.96.040404}.

\end{thebibliography}
\input{MainFile.bbl}

\end{document}